# Web Log Data Analysis by Enhanced Fuzzy C Means Clustering

V.Chitraa[1], Antony Selvadoss Thanamani[2]

[1]Assistant Professor, CMS College of Science and Commerce, Coimbatore, Tamilnadu, India
[2]Reader in Computer Science, NGM College (AUTONOMOUS), Pollachi, Coimbatore, Tamilnadu, India

## ABSTRACT

*World Wide Web is a huge repository of information and there is a tremendous increase in the volume of information daily. The number of users are also increasing day by day. To reduce users browsing time lot of research is taken place. Web Usage Mining is a type of web mining in which mining techniques are applied in log data to extract the behaviour of users. Clustering plays an important role in a broad range of applications like Web analysis, CRM, marketing, medical diagnostics, computational biology, and many others. Clustering is the grouping of similar instances or objects. The key factor for clustering is some sort of measure that can determine whether two objects are similar or dissimilar. In this paper a novel clustering method to partition user sessions into accurate clusters is discussed. The accuracy and various performance measures of the proposed algorithm shows that the proposed method is a better method for web log mining.*

## 1. INTRODUCTION

Internet has become buzzword of today's world. Ocean of information is available in internet and people use it for different purposes. Finding information in this voluminous data is a difficult task for the users. For a specific user only a small amount of web is useful and remaining are irrelevant. Systems are required in client and server side for finding out the desired information. Web mining is an active research area to extract useful information in the ocean of data. Data mining techniques are applied in web data for this purpose.

There are three categories of web mining. Web Content Mining is the process of extracting useful information from the contents of Web documents. In web structure mining the site map which depicts the structure of a typical web graph consists of web pages as nodes, and hyperlinks as edges connecting between two related pages are analyzed. Web Usage Mining is the application of data mining techniques to discover interesting usage patterns from Web data, in order to understand and better serve the needs of Web based applications. Usage data reveals the intention, identity and origin of Web users along with their browsing behavior at a Web site.

This is considered as an implicit mining since without knowledge of the users of the website the log file is analyzed. The purpose of Web usage mining is to gather useful information about navigation patterns. There are three phases of Web Usage Mining such as preprocessing, pattern discovery and analysis [13]. Preprocessing is an important phase and it is done in six steps. In the first step entries with failed status code, robot traversed entries are cleaned. The second step is user identification step where users are identified. In third step different sessions of users are segregated by using time oriented method or navigation oriented methods. Missing entries due to proxy server and cache in client systems are added in the next step. The sessions are transformed into a matrix format for further analysis. In the second phase of pattern discovery the user navigation patterns are clustered into groups. Web user clusters do not necessarily have crisp boundaries. Hence researchers interested in using the using fuzzy sets for clustering of web resources. In the third phase patterns are analyzed and matched with the incoming pattern to find out the similar user and the navigation patterns. This paper focuses on feature selection, clustering and classification of users by using vector machines. A novel soft clustering method to cluster user sessions is proposed in this paper. The proposed technique clusters the user sessions accurately and efficiently. Clusters thus formed are classified with machine learning algorithms in the pattern discovery phase. The remainder of this paper is organized as follows: Section 2 presents literature of existing works. Section 3 deals with the methodology for feature reduction using ICA, novel clustering technique, implementation of Support Vector Machines and Relevance Vector Machine to classify a new user. Experimental results are given in Section 4. At last a summary of the work is given in Section 5.

## 2. Related Work

The heterogeneous and unstructured information available in the web reduces the analysis to a greater extent. Therefore, the preprocessing phase is a prerequisite for discovering patterns and it transforms the raw click stream data into a set of user profiles [4]. Data preprocessing is a difficult task with a number of challenges. Hence this work gives rise to variety of algorithms and heuristic techniques as merging and cleaning, user and session identification etc [14]. Various research works are carried in different phases of preprocessing.

After cleaning and users are identified then sessions are created by dividing which is defined as a set of pages visited by the same user within the duration of one particular visit to a website. A user may have a single or multiple sessions during a period. Once a user has been identified, the click stream of each user is portioned into logical clusters. There are three methods in session reconstruction. There are two methods available to find sessions. One depends on time and one on navigation. Time oriented method is based on specific time browsed by user and is called as page viewing time. It varies from 25.5 minutes [1] to 24 hours [17] while 30 minutes is the default timeout by Cooley [14]. Features are the attributes of session matrix in the proposed work like individual user IP address, time etc., Feature selection phase reduces the less important attributes that are not useful for further web extraction or usage.

After sessions are discovered the similar user navigation patterns are grouped by clustering. Clusters are created by estimating the similarities and differences within a data set and group similar data. The performance of a clustering method is measured by its inter-cluster similarity and the intra-cluster similarity where the former one should be low and latter should be high.
Web user clustering involves grouping of web users. Clustering algorithms are classified into four main groups as partitioning algorithms, hierarchical algorithms, density-based algorithms, and grid-based algorithms. Partitioning algorithms divides of $N$ objects into a set of $k$ clusters such that the partition optimizes a given criterion. A hierarchical method creates hierarchical decomposition of objects and it is classified into agglomerative (bottom-up) approaches and divisive (top-down) approaches based on how the hierarchy is formed. Density-based clustering algorithms locate clusters by constructing a density function that reflects the spatial distribution of the data points.

Depending on the ability of density connected points to reach the maximal density, the density based notion of a cluster is defined. Grid-based algorithms performs all the operations in a grid structure which is formed by partitioning the space into a finite number of cells. The main advantage of the approach is its fast processing time.

However, there exist some important differences between clustering in conventional applications and clustering in web mining. Due to non-numerical nature of web data Runkler and Beadek [16] proposed relational clustering method for clustering.. Web user clusters tend to have vague or imprecise boundaries [9]. Fuzzy clustering is observed to be a suitable method to handle ambiguity in the data, since it enables the creation of overlapping clusters and introduces a degree of membership of data object in each cluster. Fuzzy modeling is proposed by Shi [16] by considering the time duration that a user spends at a URL .

A wide variety of classification methods have become available during the last three decades. Among them, the most utilized general-purpose methods can be grouped into several categories such as Inductive methods , Coverage methods , Bayesian methods, Linear methods, Non-linear methods , Lazy methods . Vector machines are observed as a most appropriate method for web user classification. Support Vector Machines are the most effective machine learning algorithm for many complex binary classification problems[11]. The SVM is a supervised classification method based on the existence of a linear separator on the data. The linear separator proposed by this method can be computed in a high dimensional space induced by a kernel. There are several algorithms to perform the SVM method as incremental SVM[2], DirectSVM [15] and SimpleSVM [18].

Recently Tipping [5] introduced another machine learning technique Relevance Vector Machine (RVM) for classification which makes probabilistic predictions and yet which retains the excellent predictive performance of the support vector machine. It also preserves the sparseness property of the SVM. Indeed, for a wide variety of test problems it actually leads to models which are dramatically sparser than the corresponding SVM, while sacrificing little if anything in the accuracy of prediction. It consists of several different types of kernels. Even though two different kernels can remain on the same location, due to the sparseness property the automatic selection of proper kernel at each location by pruning all irrelevant kernels is possible.

## 3. METHODOLOGY

Web log file is collected from web server and after cleaning, users are identified. The sessions of each user are identified by navigation oriented method. Sessions are reconstructed in the form of a matrix. Generally the presence of redundant and irrelevant attributes could mislead the analysis. The storage and processing of data increases the complexity of the analysis and degrades the accuracy of the result. Increase in attributes in the web log mining not only increases the complexity of the analysis, but also degrades the performance. So feature selection is an important step in preprocessing which enhances further analysis. To group similar users clustering is essential. To identify a new user classification is an promising research area in web log analysis.

Clustering is the process of grouping navigation patterns into groups with similar features. It divides the data into clusters in a way that similar data are placed in the same classes, while dissimilar data are place in different classes. Classification, on the other hand, a data is assigned to a predefined labeled category, if it has more features similar to that group. Both these techniques are used in this research for identifying a new user to a particular class of users with similar interest.

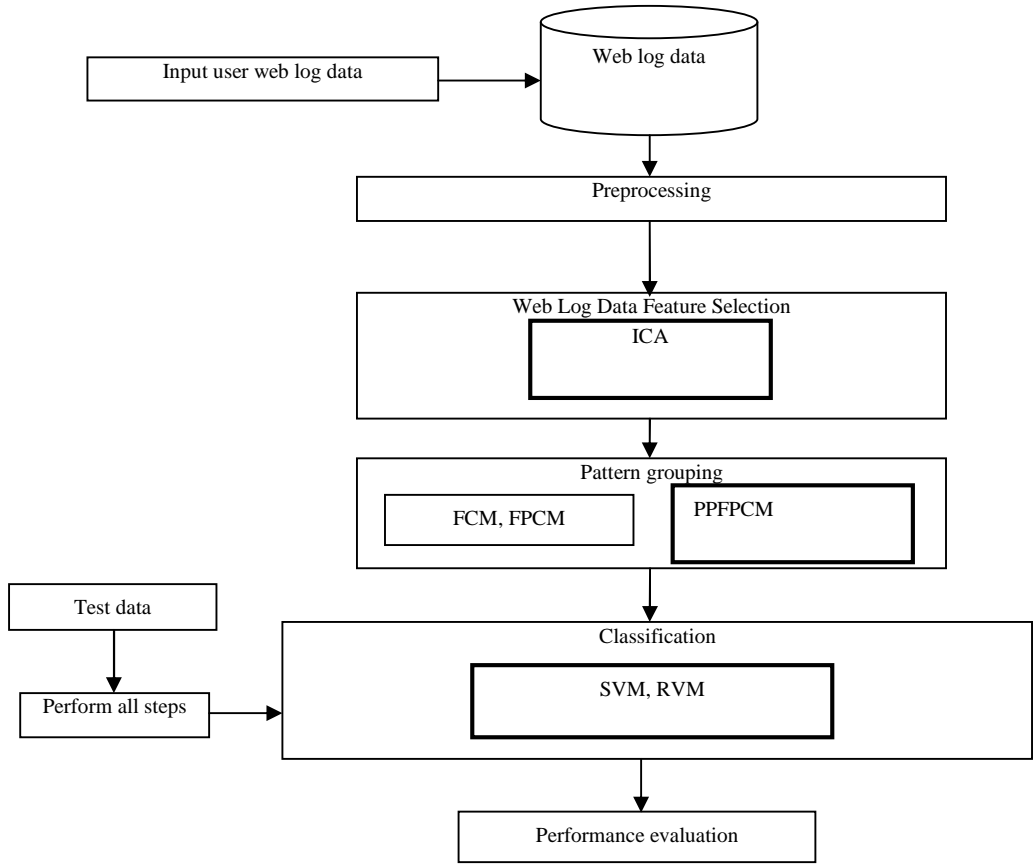

Figure 1: Architecture of entire system

## 3.1 Feature Selection by Independent Component Analysis

The presence of redundant and irrelevant attributes could mislead the analysis. For mining process if all the attributes is considered it not only increases the complexity of the analysis, but also degrades the accuracy of the result. So attributes are reduced by ICA and it is a more powerful technique to find the patterns. New patterns are selected to reduce the dimensions of matrix to enhance better performance. In the proposed method the session matrix is reduced.

In the proposed method a session matrix which consists of navigation patterns is considered. The class 'c' in this method is the unique web pages of the website browsed over a period of time. A new matrix is created after normalizing each pattern with mean and standard deviation. ICA is applied to new patterns and the minimum weighted values are deleted[1]. The remaining patterns are taken as an input for next phase. The patterns constructed are strongly related to interaction between user related variables.

The purpose of ICA is to linearly transform the original matrix data into components which are as much as statistically independent[7]. The task of ICA is to find Independent matrix $W$ to make $y$ = $Wx$ where y = $(y_1, y_2, \ldots y_{N,\ldots})^T$ is called output variable and x=$(x_1,x_2,x_3,\ldots x_N.)^T$ is called observed random variable. If $y_i$ is mutually independent, then $y_i$ is the estimated value of an independent random variable s = $(s_1,s_2,s_3\ldots s_N.)$.

Steps to reduce Features by ICA

1. User Session matrix is taken as input in which columns in which columns are the web pages and rows are users and their sessions identified from preprocessing. Each row is a navigation pattern.
2. Each user navigation patterns is normalized by $(np_i - m_i)/2\sigma_i$ where $m_i$ and $_i$ are the mean and the standard deviation of $np_i$ respectively.
3. Compute the absolute mean $a_i = \frac{1}{N+1}\sum_{j=1}^{N+1}|w_{ij}|$
4. For all $w_{ij}$ in w, if $|w_{ij}| < \alpha|$ $a_i$, then shrink $|w_{ij}|$ to zero.
5. Multiply new weight matrix W' and original user navigation pattern
6. Delete columns with values zero

## 3.2 Grouping User Sessions into Groups

Penalized Posterior probability based Fuzzy C-Means (PPPFCM) algorithm for clustering user navigation patterns data is proposed in this paper to group the user navigation patterns. After the selection of feature the user patterns are detected the group the user navigation patterns which are similar**. C**lustering is one of the methods used to group the similar user navigation patterns.

Fuzzy clustering depends on the probability of degree of belongingness of a data to clusters. The group belong to the same user navigation patterns rather than belonging completely to just one cluster. Thus, user navigation patterns points on the edge of a cluster may be in the cluster to a lesser degree than user navigation patterns in the center of cluster. With fuzzy c-means, the centroid of a cluster is the mean of all user navigation patterns points, weighted by their degree of belonging to the cluster. The navigation pattern of a user has a set of coefficients giving the degree of being in the $k^{th}$ cluster $w_k(x)$.

$$c_k = \frac{\sum_x w_k(x)^m x}{\sum_x w_k(x)^m} \qquad\qquad (1)$$

Where $w_k(x)$ is the degree of belonging which is inversely related to the distance from patterns from different sessions. It also depends on a parameter m that controls how much weight is given to the closest center of the user navigation patterns. For the initial cluster centroids for each user navigation patterns, fuzzification parameters are considered. The membership values are calculated using objective function.

Penalized Fuzzy C-Means algorithm for clustering user navigation patterns data is considered in this work [11]. In traditional Fuzzy C-Means algorithm the neighborhood user navigations are not incorporated. In order to incorporate the spatial context, the objective function of Fuzzy C-Means algorithm is to be modified by introducing penalty factor to produce more meaningful fuzzy clusters. The cluster centroids for each user navigation patterns is to be computed and membership values will be calculated.

In Penalized FCM clustering algorithm, each user session is not exactly analyzed to evaluate membership value. By using Bayesian statistics the posterior probability is calculated by the probability of the parameters which contrasts with the likelihood function when given with the evidence. They are considerably higher than equivalent nonparametric methods and erroneous conclusions will be made more often. The method will assign user navigation patterns to the class with highest membership.

**Steps for the Penalized Posterior probability based Fuzzy C-Means**

Set the cluster centroids $v_i$, fuzzification parameter q, the cluster index l and number of clusters c. Repeat until

1. Calculate membership values using Equation.

$$\mu_{ik}^* = \frac{1}{\sum_{l=1}^{c} \left( \frac{d^2(x_k, v_i) + \gamma \sum_{j=1}^{N} (1-\mu_{ij})^q w_{kj}}{d^2(x_k, v_l) + \gamma \sum_{j=1}^{N} (1-\mu_{ij})^q w_{kj}} \right)^{\frac{1}{(q-1)}}} \qquad (2)$$

2. Evaluate the highest membership value result $\mu_{ik}^*$ based on the probability function .The posterior probability is the probability of the parameters $\mu_{ik}^*$ given the evidence $(\mu_{ik}^* | x_k)$ . $P(\mu_{ik}^* | x_k)$ is the probability of the evidence given by the parameters and it is different from the likelihood function in conventional FCM . The posterior probability is defined as

$$P(\mu_{ik}^* | x_k) = \frac{P(x_k | \mu_{ik}^*) P(\mu_{ik}^*)}{p(x_k)} \qquad (3)$$

3. Compute the cluster centroids for user navigation patterns $np_i$ using Eq.

$$v_i^* = \frac{\sum_{k=1}^{N} (P(\mu_{ik}^* | x_k))^q x_k}{\sum_{k=1}^{N} (P(\mu_{ik}^* | x_k))^q} \qquad (4)$$

Until convergence criteria of the similar user navigation patterns are clustered.

After convergence a defuzzification process takes place, in order to convert the user navigation patterns matrix to a crisp partition. Among many methods available for defuzzification the maximum membership procedure is the best method for user navigation patterns. The procedure assigns the user navigation patterns k to the class C with the highest membership

$$C_k = \arg_i \left( \max \left( P(\mu_{ik}^* | x_k) \right) \right), i = 1, \dots c \qquad (5)$$

## 3.3 Classification

Classification is the process of classifying a data item into one of several predefined classes. After similar navigation patterns are clustered from above clustering methods then next step is to classify user navigation patterns to a class and when new user navigation patterns entered into web page similar user navigation patterns are identified and classified using classification methods. In this work proposed vector methods such as SVM, RVM are used for classification.

### 3.3.1 Classification by Support Vector Machines

Support Vector Machines based on Structural Risk Minimization acts as one of the best approach for classification. The principal behind this theory is to learn from finite data and controls generalization ability of learning machines which reduces the error. The data is separated by a hyper plane in the n-dimensional space. The data which are closest to the hyperplane are called support vectors which acts as representatives. The hyperplane that separates two groups of data is expressed in the objective function

$$w.x + b = 0 \qquad\qquad (6)$$

where x is a set of training vectors, w represent vectors perpendicular to the separating hyperplane and b represents the offset parameters which allows increase of the margin.
For n-dimensional data n-1 hyperplanes are to be introduced. A positive slack variable $\epsilon$ is introduced in the objective function in order to add some flexibility in separating the categories[ ]. The improvised objective function is

$$y_i(w.x_i+b) >= 1- \epsilon_i \qquad\qquad (7)$$

where $\epsilon >= 0$. For non-linear classification kernel functions are used instead of every dot product. It maps the input space into a high dimensional space through a non-linear transformation. The kernel mapping provides a common base for most of the commonly employed model architectures, enabling comparisons to be performed [8]. There are several kernel functions like Gaussian, Laplace, Radial Basis Function are available in which Laplacian is observed as suitable for the log data classification.

$$k(x,y) = \exp\left(-\frac{||x-y||}{\sigma}\right) \qquad\qquad (9)$$

Given a training set of instance-label pairs $(x_i; y_i)$; $i = 1… l$ where $x_i \in Rn$ and $y \in \{1,-1\}^l$, then the support vector machines requires the solution for the objective function

$$y_i(w.\quad(x_i) +b) >= 1- \epsilon_i \qquad\qquad (10)$$

The function maps training vectors xi into a higher dimensional space. A linear separating hyperplane with the maximal margin in this higher dimensional space is found by SVM.

The steps to train using SVM Classifier is
1. Prepare and normalize the cluster data
2. Consider the Laplacian Kernel
3. Use Cross-validation to find the best parameters and train the whole training data set.
4. Test the data and find accuracy.

The advantages of SVM are

- Effective in high dimensional space
- Memory efficient
- Accuracy.

There are few drawbacks such as

- Non-probabilistic classification which is a need for user behavior classification
- An increase in number of support vectors with the size of increasing training set.

Due to these reasons the computational complexity increases and experimentation is done with another technique which overcomes these pitfalls.

### 3.3.2 Classification By Relevance Vector Machines

There are certain drawbacks observed in SVM like complexities increase with increase in classes and kernel functions are centered on training data points which must be positive definite. So another machine learning technique Relevance Vector Machines is considered for classification. RVM rely on Bayesian inference learning. RVM has the highest capacity to find exact user patterns and probabilistic solutions.

### Steps in RVM for Classification

Step 1 Initialize parameters and select Laplacian Kernel function to design matrix
Step 2. Establish a suitable convergence criteria for   and
Step 3. Fix a threshold value $_{Th}$ which it is assumed an $_i$ is tending to infinity upon reaching it.
Step 4. Initialize   and
Step 5. Calculate m = $^T$t where  =(A + $^T$ )$^{-1}$
Step 6. Update $_i = \gamma i$ / mi$^2$ and  =(N - $_i \gamma i$) / || t-  m||
Step 7. Apply pruning to $_i$ and basis functions $_{i> Th}$
The  steps 5 to 7 until convergence criteria is met. Whenever a new user arrives after doing preprocessing, the resultant pattern is estimated with a target value 't' for a new input x' by using the objective function

t=m$^T$  (x')                                              (8)

# 4. Experimental Results

The proposed web log preprocessing is evaluated  in this section.  The results of different stages are given as follows. The method is implemented by using  MATLAB.

## 4.1 Web log data

The weblog data considered for evaluation is collected from a furniture shop web server during the period of May to August, 2013.  Initially the log file consists of   23141  raw log entries with noisy entries like gif, jpeg etc which are not necessary for web log mining.

## 4.2 Data Cleaning

Data cleaning phase is performed and irrelevant entries are removed.  Entries with graphics and videos format such as gif, JPEG, etc., and  entries with robots traversal are removed. The number of records resulted after cleaning phase is 19687 and it is represented in figure 2.

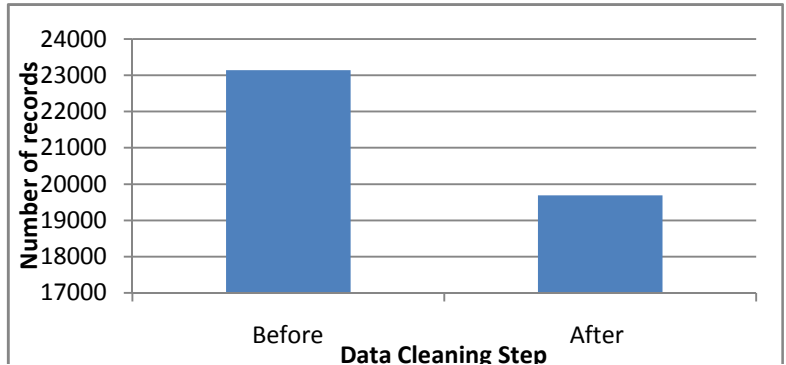

Figure 2: No of log Entries before and after Data Cleaning

## 4.3 Session Matrix Construction

There are 358 unique users identified after applying the algorithm. Session identification process is carried out. Sessions are reconstructed in a matrix format which gives details about the frequency and navigation pattern of different users. A part of the result obtained by using the session identification process is represented in matrix format is represented in figure 3. U1, U2, etc., represents the user with particular IP address. S1, S2 etc., gives the session constructed as per proposed method. The columns 1,2,3….are the web pages browsed by users in the respected sessions, As presented in the figure, it can be seen that for user 1 there are two sessions in which first session has the URL traversal 1-2. For user 2, only one session is resulted i.e., 5-6-7---.

| Users | Sessions | 1 | 2 | 3 | 4 | 5 | 6 | 7 |
|-------|----------|---|---|----|---|-----|---|---|
| U1 | S1 | 1 | 2 | 10 | | | | |
| U1 | S2 | 2 | 0 | 0 | 0 | 0 | 0 | 2 |
| U2 | S1 | 0 | 0 | 0 | 0 | 2 | 1 | 1 |
| U3 | S1 | 0 | 1 | 0 | 2 | 100 | | |
| U3 | S2 | 0 | 0 | 0 | 0 | 0 | 0 | 1 |
| U4 | S1 | 0 | 0 | 0 | 2 | 1 | 0 | 0 |
| U5 | S1 | 0 | 4 | 0 | 0 | 0 | 1 | |

Figure 3.Sample User Session Matrix

## 4.4 Feature Reduction by ICA.

In the given input data the matrix constructed consists of 14 columns in which most of the columns are irrelevant and after applying there are only 5 columns which enhances further analysis in terms of time and complexity.

## 4.5 Clustering By PPPFCM

The sessions are divided into 14 clusters. User-Sessions are distributed among these 14 clusters. A sample of 3 clusters is given in Figure 4.

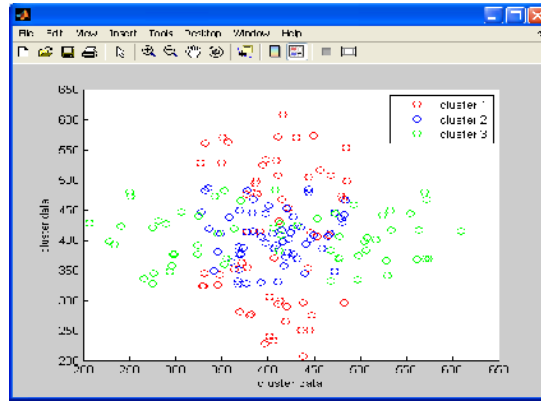

Figure 4.Sample Clusters

To measure the clustering accuracy of proposed system Penalized Posterior Probability FCM and existing methods FCM(Fuzzy C means) clustering, Penalised FCM clustering, the parameters such as Rand Index, Sum of Squared Error (SSE) are used.

### 4.4.1 Rand Index

The Rand index is a measure used to compare induced clustering structure$(C_1)$ and a clustering structure $(C_2)$. Let a be the number of instances that are assigned to the cluster in $C_1$ and $C_2$. b be the number of instances that are in the same cluster $C_1$, but not in the cluster in $C_2$. 'c' be the number of instances that are in the same cluster in $C_2$, but not in the cluster in $C_1$, and 'd' be the number of instances that are assigned to different clusters in $C_1$ and $C_2$. The quantities a and d can be interpreted as agreements, and b and c as disagreements. The Rand index is defined as:

$$Rand\ Index = \frac{a+b}{a+b+c+d} \qquad\qquad (9)$$

The Rand index lies between 0 and 1. When the two partitions agree perfectly, the Rand index is 1.The clustering accuracy is high and the rand index value is nearly equal to one or else the accuracy of clustering results is less. The clustering accuracy results of methods is shown in Figure 5. It shows proposed PPPFCM have higher rand index rate than FCM and PFCM methods.The results of clustering methods for Rand Index is shown in table 1

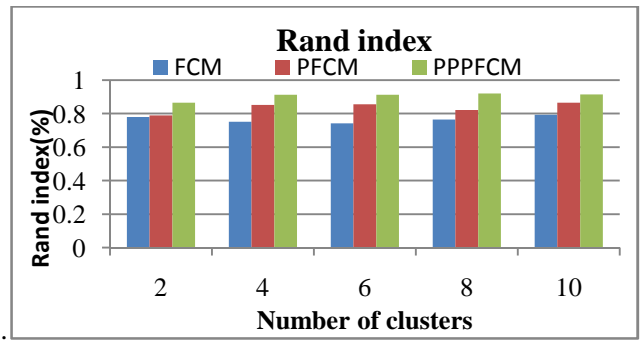

Figure 5. Rand index of clustering methods

Table 1. Rand index of clustering results

| Number of clusters | Rand index (RI) (%) | | |
|---|---|---|---|
| | FCM | PFCM | PPPFCM |
| 2 | 0.78 | 0.79 | 0.8654 |
| 4 | 0.7512 | 0.852 | 0.9123 |
| 6 | 0.7412 | 0.8562 | 0.9125 |
| 8 | 0.7652 | 0.8215 | 0.921 |
| 10 | 0.7923 | 0.8645 | 0.915 |

### 4.4.2 Sum of Squared Error (SSE)

SSE is the simplest and most widely used criterion measure for clustering. It measures the compactness of the clusters. It is calculated as

$$SSE = \sum_{k=1}^{K} \sum_{x_i \in C_k} ||x_i - \mu_k||^2 \tag{10}$$

where $C_k$ is the set of instances in cluster $k$, $\mu_k$ is the vector mean of cluster $k$. The components of $\mu_k$ are calculated as:

$$\mu_{k,j} = \frac{1}{N_k} \sum_{x_i \in C_k} x_{i,j} \tag{11}$$

where $N_k = |C_k|$ is the number of instances belonging to cluster $k$. The SSE error values is less below to 0.01 ,error values of clustering results are less .The results of SSE among clustering methods for user session matrix is shown in Figure 6. It shows that proposed system have less error value when compare to FCM and PFCM methods, the results are also tabulated in table 3 .

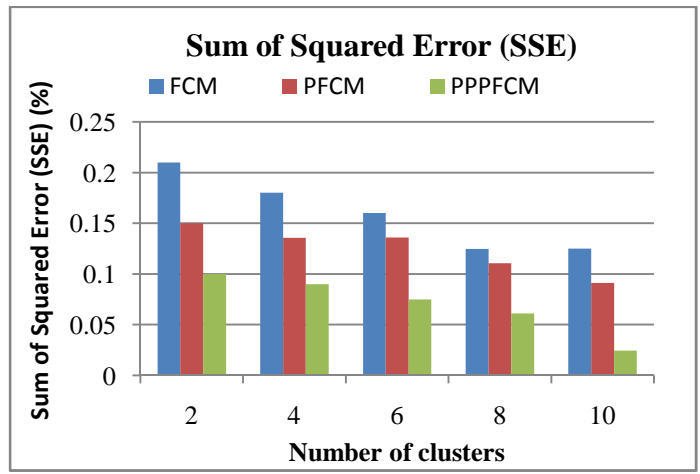

Figure 6. Sum of Squared Error (SSE) of clustering methods

Table  3: Sum of Squared Error (SSE) of clustering methods

| Number of clusters | Sum squared Error (SSE)(%) | | |
|---|---|---|---|
| | FCM | PFCM | PPPFCM |
| 2 | 0.21 | 0.15 | 0.1 |
| 4 | 0.18 | 0.1358 | 0.09 |
| 6 | 0.16 | 0.1359 | 0.075 |
| 8 | 0.1248 | 0.1105 | 0.0612 |
| 10 | 0.1251 | 0.0912 | 0.02458 |

### 4.4.3. F-measure

All three algorithms are compared based on the  effectiveness of the clustering results. The F-measure is calculated by combining  precision and recall values of clusters. Let us consider $c_i$ designate the number of session  in class $i$, and $c_j$ designate the number of  session in cluster j. Moreover, let $C_{ij}$ designate the number of items of class i present in cluster j. The  $prec(i, j)$ is the precision of cluster j with respect to class i and $rec(i, j)$  is the recall of a cluster $j$ with respect to class $i$ as

$$prec(i, j) = \frac{c_{ij}}{c_j} \qquad\qquad (12)$$

$$rec(i, j) = \frac{c_{ij}}{c_i} \qquad\qquad (13)$$

The f-measure, $F(i, j)$, of a class $i$ with respect to cluster $j$ is then defined as

$$F(i, j) = \frac{2 \ (prec(i,j) \cdot rec(i,j))}{prec(i,j) + rec(i,j)} \qquad\qquad (14)$$

The values of F-Measure is high when formation of cluster results are correct. Proposed PPPFCM clustering algorithm have higher clustering accuracy  than existing methods such as FCM and PFCM methods. Results are tabulated in table 4 and comparison of performance is  is shown in Figure 6.

Table 4.F-measure of clustering methods

| Number of clusters | F-measure (%) | | |
|---|---|---|---|
| | FCM | PFCM | PPPFCM |
| 2 | 0.71 | 0.725 | 0.9125 |
| 4 | 0.723 | 0.741 | 0.9189 |
| 6 | 0.7324 | 0.7512 | 0.9215 |
| 8 | 0.7413 | 0.7648 | 0.9289 |
| 10 | 0.7498 | 0.7712 | 0.9315 |

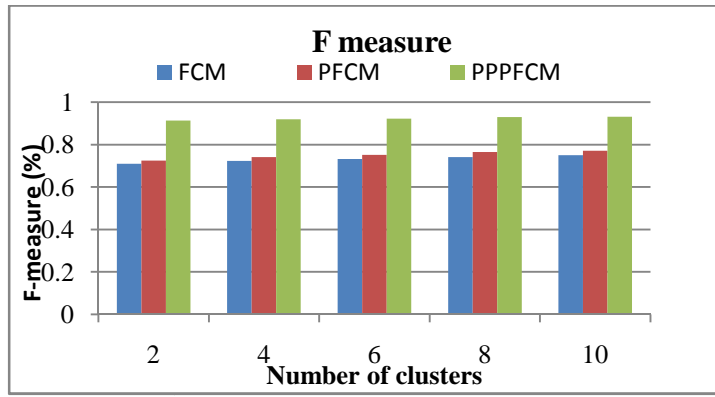

Figure 6: F-measure of clustering methods

## 4.5 Classification By Support Vector Machines And Relevance Vector Machines

Classification is done by implementing both vector machine methods. The performance measures of these methods are measured by various parameters. The true positive rate TPrate = TP/(TP+FN) represents the rate of recognition of the positive class. It is also known as sensitivity. The corresponding measurement for the negative class is the true negative rate (TNrate), also called specificity, and is computed as the number of negative examples correctly identified, out of all negative samples TNrate=TN/(TN+FP). The results for the measures calculated after classification by SVM and RVM are tabulated in table 5 and a graphic representation is shown in figure 7 below.

Table 5. Measures of SVM and RVM Classification

| Measures | SVM | RVM |
|---|---|---|
| Sensitivity | 0.4077 | 0.6035 |
| Specificity | 0.2397 | 0.2036 |

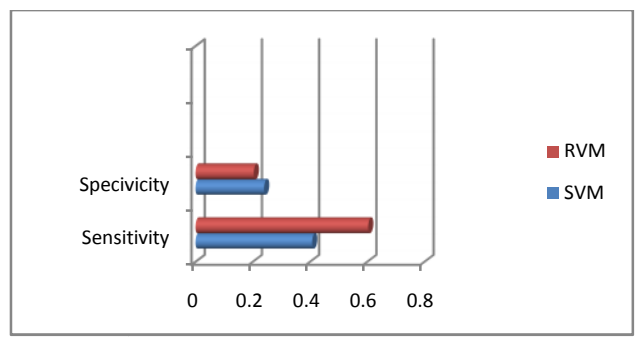

Figure 7. Comparison of SVM and RVM Classification

The most widely employed such metric is the accuracy (Acc) of the classifier, or its complement – the error rate (Err). Acc is defined as the ratio between the total number of correctly classified instances and the total number of instances, Acc = (TP+TN)/(TP+TN+FP+FN), They are usually expressed as percentages. A comparison of accuracy of both the methods are shown in Figure 8.

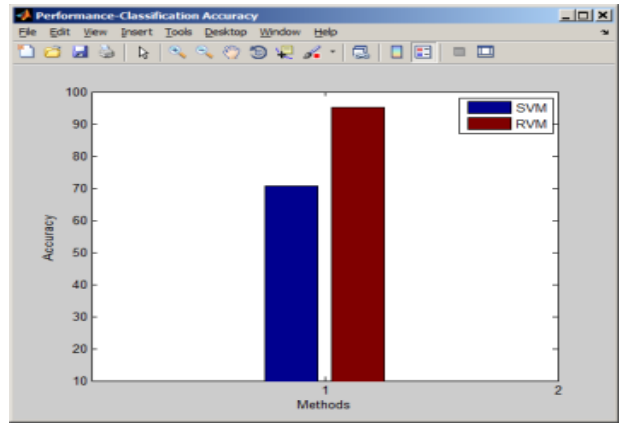

Figure 8.Accuray of SVM and RVM

# 5. CONCLUSION

In this paper, a novel approach based on Fuzzy C Means in fuzzy environments is proposed to cluster the web user transactions. This approach is groups the similar user navigation patterns. The algorithm enhances the FCM, and Penalized FCM clustering algorithm by adding Posterior Probability to find highest membership for a member to add in a cluster. Classification is carried out by SVM and RVM for classifying a new user to a particular group. The method is experimented and evaluated and found it is better method for clustering than the existing methods.

## AUTHORS

**Mrs. V. Chitraa** is a doctoral student in Manonmaniam Sundaranar University, Tirunelveli, Tamilnadu. She is working as an Assistant Professor in CMS college of Science and Commerce, Coimbatore and a member of IEEE. Her research interest lies in Web Mining, Database Concepts and knowledge mining. She has published many papers in reputed international journals and conferences.

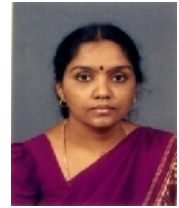

**Dr. Antony Selvadoss Thanamani** is working as a Reader in the department of Computer Science in NGM college with a teaching experience of about 26 years.His research interests include Knowledge Management, Web Mining, Networks,Mobile Computing, and Telecommunication. He has guided more than 41 M.Phil.Scholars and many Ph.D., scholars. He has presented more than 50 papers. He has attended more than 15 workshops, seminars and published about 8 books and 16 papers

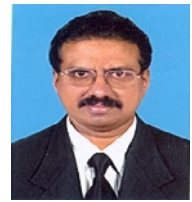